\documentclass[%
 reprint,
 superscriptaddress,
 bibnotes,
 amsmath,amssymb,
 aps,
 prb,
 floatfix
]{revtex4-2}

\usepackage[dvipdfmx]{graphicx}
\usepackage{braket}
\usepackage{dcolumn}
\usepackage{bm}
\usepackage{hyperref}
\usepackage{float}
\hypersetup{
  colorlinks   = true,
  urlcolor     = black,
  linkcolor    = black,
  citecolor   = blue
}
\usepackage{etoolbox}
\apptocmd{\sloppy}{\hbadness 10000\relax}{}{}

\begin{document}

\title{Evolution of crystal field and intraionic interactions in the ilmenite $A$IrO$_3$ ($A$ = Mg, Zn, Cd) and hyperhoneycomb $\beta$-ZnIrO$_3$}

\author{Yuya Haraguchi}
\email[]{chiyuya3@go.tuat.ac.jp}
\affiliation{Department of Applied Physics and Chemical Engineering, Tokyo University of Agriculture and Technology, Koganei, Tokyo 184-8588, Japan}

\author{Hiroko Aruga Katori}
\affiliation{Department of Applied Physics and Chemical Engineering, Tokyo University of Agriculture and Technology, Koganei, Tokyo 184-8588, Japan}

\author{Kenji Ishii}
\affiliation{Synchrotron Radiation Research Center, National Institutes for Quantum and Radiological Science and Technology, Hyogo 679-5148, Japan}

\author{Hakuto Suzuki}
\email[]{hakuto.suzuki@tohoku.ac.jp}
\affiliation{Frontier Research Institute for Interdisciplinary Sciences, Tohoku University, Sendai 980-8578, Japan}
\affiliation{Institute of Multidisciplinary Research for Advanced Materials (IMRAM), Tohoku University, Sendai 980-8577, Japan}
\date{\today}

\begin{abstract}

Spin-orbit Mott insulators with the $t_{2g}^5$ electron configuration are promising platforms for the Kitaev spin liquid, yet fine-tuning of their crystal structures is essential to suppress non-Kitaev interactions. Here, we investigate the local electronic structures of the ilmenite iridates $A\mathrm{IrO}_3$ ($A = \mathrm{Mg}, \mathrm{Zn}, \mathrm{Cd}$) and the hyperhoneycomb $\beta\text{-}\mathrm{ZnIrO}_3$ using Ir $L_3$-edge resonant inelastic x-ray scattering (RIXS). Multiplet analysis of the RIXS spectra reveals a systematic evolution of the crystal field and intraionic interaction parameters upon chemical substitution at the $A$-site. We observe an enhancement of the trigonal distortion with increasing $A$-site ionic radius. This provides a microscopic explanation for the deviation from the ideal $J=1/2$ state and the antiferromagnetic interactions identified in $\mathrm{CdIrO}_3$. Furthermore, the local multiplet parameters of ilmenite $\mathrm{ZnIrO}_3$ and hyperhoneycomb $\beta\text{-}\mathrm{ZnIrO}_3$ are found to be nearly identical, demonstrating that their different magnetic ground states are primarily governed by their distinct lattice structures rather than the single-ion properties. These findings establish a solid foundation for understanding how local crystal-field distortions control the magnetic Hamiltonian in Kitaev candidate materials.
\end{abstract}

\maketitle

\section{Introduction}
The exactly soluble Kitaev honeycomb model \cite{Kitaev.A_etal.Ann.-Phys.2006} provides a canonical example of quantum spin liquids \cite{Savary.L_etal.Rep.-Prog.-Phys.2017}. This model also holds promise as a basis for fault-tolerant quantum computation by utilizing the Majorana fermions emergent from the fractionalization of the local spin $S=1/2$. The Kitaev model has been implemented in a quantum material platform, after a theoretical proposal by Jackeli and Khalliulin \cite{Jackeli.G_etal.Phys.-Rev.-Lett.2009}, which suggests that spin-orbit Mott insulators with edge-shared octahedral crystal field arrangement can host bond-dependent interactions between the $J=1/2$ pseudospins. This proposal has driven the search for 4$d$ and 5$d$ transition metal compounds with the honeycomb lattice structure that do not show magnetic phase transitions down to low temperatures \cite{Rau.J_etal.Annu.-Rev.-Condens.-Matter-Phys.2016,Hermanns.M_etal.Annu.-Rev.-Condens.-Matter-Phys.2018,Takagi.H_etal.Nat.-Rev.-Phys.2019,Motome.Y_etal.J.-Phys.-Condens.-Matter2020}. 

\begin{figure}[ht]
  \centering
  \includegraphics[width=\linewidth]{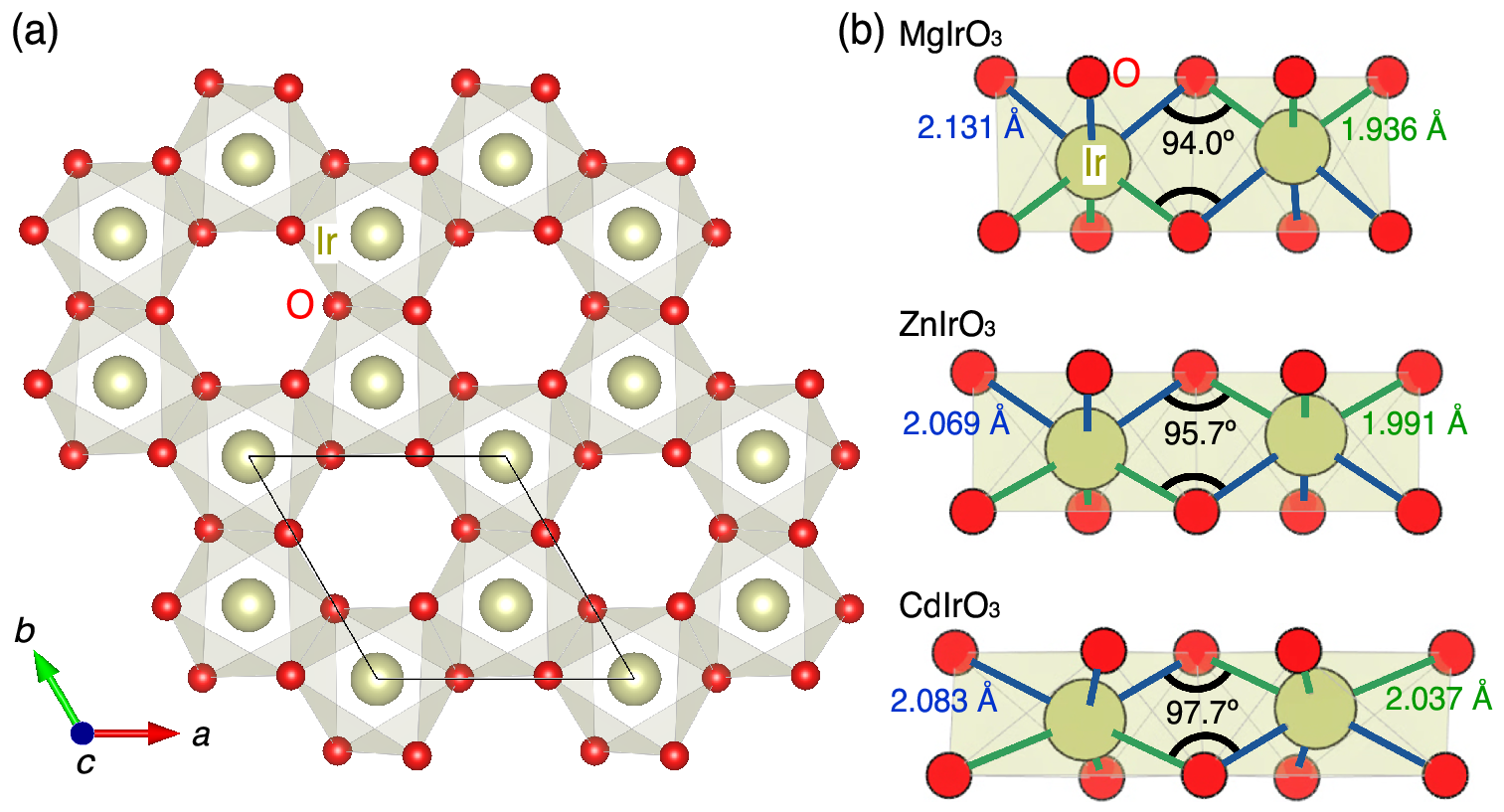}
  \caption{(a) IrO$_3$ honeycomb layers in the ilmenite iridates $A$IrO$_3$ ($A$ = Mg, Zn, Cd). The iridium and oxygen ions are shown in green and red, respectively. The figure is drawn using VESTA software \cite{Momma.K_etal.J.-Appl.-Cryst.2011}. (b) Illustration of the trigonal crystal field distortion in $A$IrO$_3$. The two types of Ir-O bonds with different lengths are shown in blue and green. The Ir-O-Ir bond angles are also indicated, showing the gradual deviation from the octahedral crystal field symmetry around the Ir ions.}
  \label{crystal}
\end{figure}

Recently, ilmenite-type 5$d$ iridium oxides $A$IrO$_3$ ($A$ = Mg, Zn, Cd) with honeycomb structures have been synthesized \cite{Haraguchi.Y_etal.Phys.-Rev.-Materials2018,Haraguchi.Y_etal.Phys.-Rev.-Materials2020}. As shown in Fig. \ref{crystal}(a), the edge-shared IrO$_6$ octahedra 
form a honeycomb lattice in the $ab$ plane, while the $A$-site ions are located between the IrO$_3$ layers. The $A$-site ions are divalent cations, such as Mg$^{2+}$, Zn$^{2+}$, and Cd$^{2+}$. The crystal field environment of the Ir$^{4+}$ ion is distorted from the cubic symmetry to trigonal symmetry due to the different Ir-O bond lengths and Ir-O-Ir bond angles, as illustrated in Fig. \ref{crystal}(b). Unlike many other monoclinic Kitaev candidate magnets, including Na$_2$IrO$_3$ \cite{Singh.Y_etal.Phys.-Rev.-B2010,Hwan-Chun.S_etal.Nat.-Phys.2015}, Li$_2$IrO$_3$ \cite{Singh.Y_etal.Phys.-Rev.-Lett.2012}, Cu$_2$IrO$_3$ \cite{Choi.Y_etal.Phys.-Rev.-Lett.2019,Haraguchi.Y_etal.J.-Phys.-Condens.-Matter2024}, H$_3$LiIr$_2$O$_6$ \cite{Kitagawa.K_etal.Nature2018}, Ag$_3$LiIr$_2$O$_6$ \cite{Bahrami.F_etal.Phys.-Rev.-Lett.2019}, and $\alpha$-RuCl$_3$ \cite{Plumb.K_etal.Phys.-Rev.-B2014,Sears.J_etal.Phys.-Rev.-B2015}, the ilmenite $A$IrO$_3$ allows the systematic control of the trigonal field via chemical substitution. 

Indeed, the antiferromagnetic order occurs at $T_N$ = 31.8 K, 46.6 K, and 90.9 K for $A$ = Mg, Zn, and Cd, respectively \cite{Haraguchi.Y_etal.Phys.-Rev.-Materials2018,Haraguchi.Y_etal.Phys.-Rev.-Materials2020}, suggesting a systematic evolution of the magnetic Hamiltonian. In MgIrO$_3$ and ZnIrO$_3$, an effective magnetic moment close to $J = 1/2$ is observed \cite{Haraguchi.Y_etal.Phys.-Rev.-Materials2018}. Furthermore, \textit{ab initio} calculations for MgIrO$_3$ and ZnIrO$_3$ \cite{Jang.S_etal.Phys.-Rev.-Materials2021} show that the bands described by the $J = 1/2$ pseudospin have a gap across the Fermi level, indicating the realization of spin-orbit Mott insulators. On the other hand, the effective moment of 2.26 $\mu_B$ is estimated in CdIrO$_3$ \cite{Haraguchi.Y_etal.Phys.-Rev.-Materials2020}, which is significantly larger than 1.73 $\mu_B$ of $J=1/2$ pseudospin. This indicates a large deviation of the local crystal field from the cubic symmetry. Additionally, CdIrO$_3$ exhibits a substantial antiferromagnetic interaction with a Weiss temperature of $\theta_{W}=-280$ K. This suggests that the distortion of the IrO$_6$ octahedra drives the modifications in the magnetic Hamiltonian. Therefore, it is imperative to precisely determine the multiplet parameters in $A\text{IrO}_3$ to reveal their impact on the resulting magnetic properties.

\begin{figure}[ht]
  \centering
  \includegraphics[width=7cm]{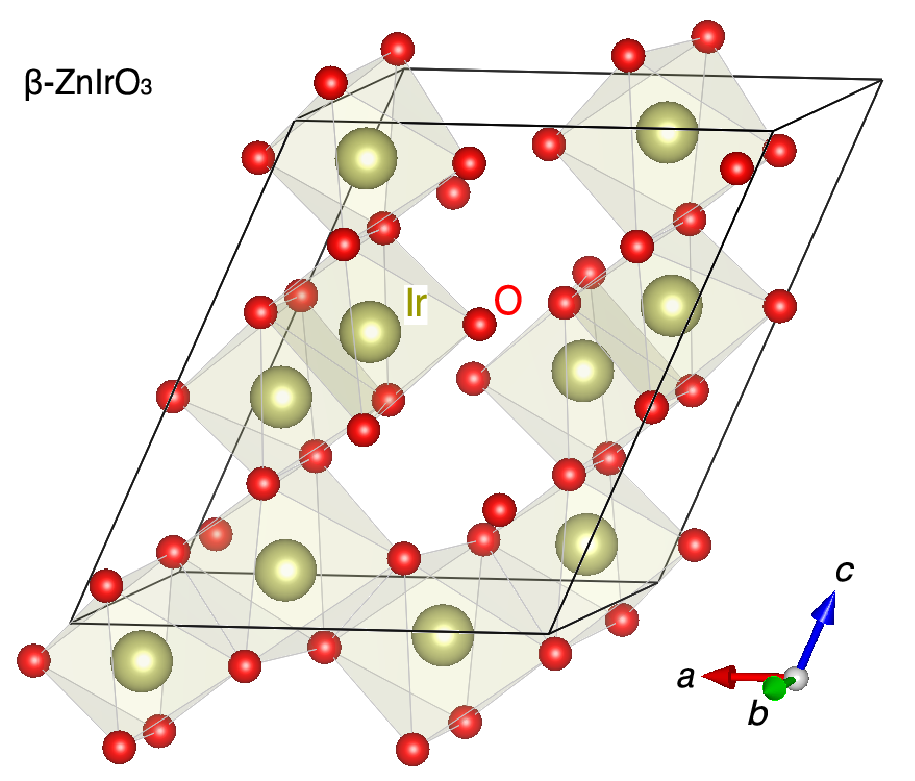}
  \caption{Hyperhoneycomb network formed by the IrO$_6$ octahedra in $\beta$-ZnIrO$_3$. The Zn ions are not illustrated.}.
  \label{hyper}
\end{figure}

Furthermore, the hyperhoneycomb iridate $\beta$-ZnIrO$_3$ has been synthesized via a topochemical metathesis reaction \cite{Haraguchi.Y_etal.Phys.-Rev.-Materials2022, Haraguchi.Y_etal.Chem.-Lett.2023}. In this compound, $\text{Ir}^{4+}$ ions form a three-dimensional hyperhoneycomb lattice (Fig. \ref{hyper}), which provides a potential platform for the three-dimensional Kitaev model. Unlike the ilmenite $A\text{IrO}_3$ phases, $\beta$-ZnIrO$_3$ is characterized by dominant ferromagnetic interactions with a positive Curie-Weiss temperature of $\theta_W \sim 45$ K. The material exhibits no long-range magnetic order down to 2 K and shows gapless excitations, as evidenced by the linear temperature dependence of the heat capacity \cite{Haraguchi.Y_etal.Phys.-Rev.-Materials2022}. These features suggest the realization of a quantum paramagnetic state proximate to a spin liquid. To provide a solid foundation for these properties, it is essential to establish the detailed properties of the $J=1/2$ pseudospins.   

In the present work, we systematically investigate the multiplet structures in polycrystalline samples of the ilmenite iridates $A$IrO$_3$ ($A$ = Mg, Zn, Cd) and the hyperhoneycomb $\beta$-ZnIrO$_3$ using resonant inelastic x-ray scattering (RIXS) \cite{Ament.L_etal.Rev.-Mod.-Phys.2011,Ishii.K_etal.J.-Phys.-Soc.-Jpn.2013} at the Ir $L_3$ edge. We reveal a systematic evolution of the multiplet parameters and crystal field environment across these materials. These results provide microscopic foundations for how crystal field distortions govern the magnetic properties of these iridates.

RIXS experiments were conducted using the hard x-ray RIXS spectrometer at the BL11XU of SPring-8 \cite{Ishii.K_etal.J.-Electron.-Spectrosc.2013}. The incident x-ray energy was tuned to the Ir $L_3$ absorption edge ($\sim$ 11.214 keV). The incident x-ray photons were $\pi$-polarized. Scattered photons were collected at a fixed scattering angle of 90$^\circ$ using a $\mathrm{Si}$ ($884$) diced spherical analyzer and a single photon counting micro strip detector. The polarizations of the outgoing photons were not analyzed; thus, the RIXS intensity comprises the $\pi$-$\pi$ and $\pi$-$\sigma$ scattering channels. The incident x-ray beam was focused to a beam spot of 34$\times$600 $\mu$m$^{2}$ (H $\times$ V).  All measurements were performed at $T=300$ K. The overall energy resolution, defined as the full width at half maximum of the nonresonant spectrum from Kapton tape, was 46 meV.

\section{Results and discussion}

Figure \ref{fig:rixs}(a) shows the x-ray absorption spectra (XAS) of $A$IrO$_3$ and $\beta$-ZnIrO$_3$. All the XAS spectra exhibit a main $L_3$ peak at 11.217 keV (vertical dashed line), corresponding to the electronic transition from the Ir $2p_{3/2}$ to the $e_g$ orbitals of the Ir $5d$ states. The lineshapes are nearly identical, confirming that the Ir ions in these compounds are in the $4+$ valence state. The slight intensity reduction in ZnIrO$_3$ and $\beta$-ZnIrO$_3$ is ascribed to the self-absorption of the emitted x-rays by the Zn$^{2+}$ ions. For the subsequent Ir $L_3$-edge RIXS measurements, the incident photon energy was tuned to 11.214 keV (vertical purple line), which is 3 eV below the XAS peak and approximately corresponds to the transition to the $t_{2g}$ orbitals of the Ir $5d$ states.

Figure \ref{fig:rixs}(b) presents the Ir $L_3$-edge RIXS spectra of $A$IrO$_3$ and $\beta$-ZnIrO$_3$. In addition to the quasi-elastic excitations within the $J=1/2$ sector, one identifies spin-orbit excitations to the $J = 3/2$ states (indicated by vertical bars) and crystal field transitions to the $t_{2g}^4e_g$ multiplets (indicated by circles and triangles). The multiplet structures are analogous to those in other iridium \cite{Gretarsson.H_etal.Phys.-Rev.-Lett.2013} and ruthenium-based Kitaev magnets \cite{Suzuki.H_etal.Nat.-Commun.2021,Lebert.B_etal.Phys.-Rev.-B2023,Gretarsson.H_etal.Phys.-Rev.-B2024}. These multiplet structures firmly establish the formation of $J=1/2$ pseudospins in $A$IrO$_3$ and $\beta$-ZnIrO$_3$. The nearly identical peak energies in ZnIrO$_3$ and $\beta$-ZnIrO$_3$ indicate the nearly identical local crystal field environment and intraionic Coulomb interactions among the Ir $5d$ electrons in these compounds. In contrast, the ilmenite series $A$IrO$_3$ exhibits a systematic shift in peak energies, reflecting the evolution of the crystal field and effective strengths of Coulomb interactions with different $A$ ions.

To qualitatively extract the multiplet parameters in these iridates from the RIXS data, we have performed multiplet calculations of the RIXS intensity. The analysis procedure is analogous to those in Refs. \cite{Suzuki.H_etal.Nat.-Commun.2021,Takahashi.H_etal.Phys.-Rev.-Lett.2021,Gretarsson.H_etal.Phys.-Rev.-B2024}. Specifically, we have diagonalized the ionic model Hamiltonian for the $d^5$ electron configuration, comprising the intra-atomic Coulomb interaction 
$H_\mathrm{C}$ \cite{Sugano.S_etal.1970,Georges.A_etal.Annu.-Rev.-Condens.-Matter-Phys.2013}, the intra-atomic SOC $H_\mathrm{SOC}$, the cubic crystal field $H_\mathrm{cub}$, and the trigonal crystal field $H_\mathrm{trig}$:

\begin{figure}[ht]
  \centering
  \includegraphics[width=\linewidth]{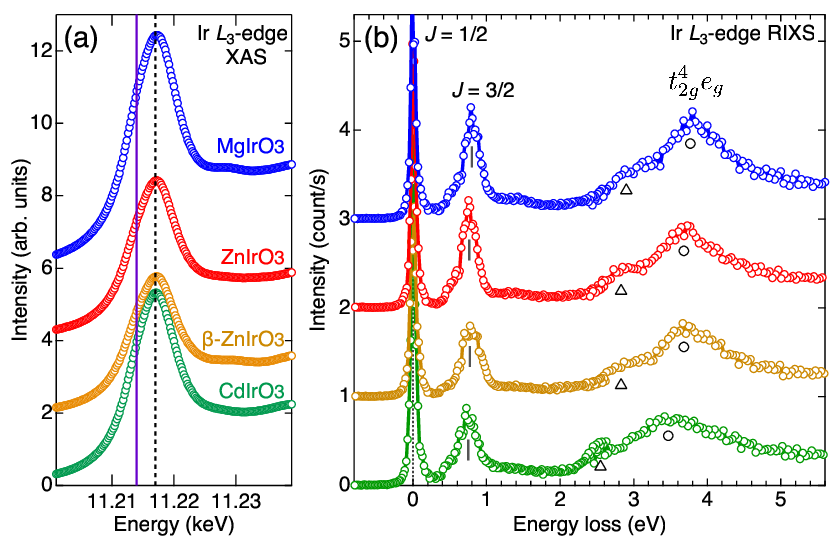}
  \caption{(a) Ir $L_3$-edge x-ray absorption spectra (XAS) of the ilmenite iridates $A$IrO$_3$ ($A$ = Mg, Zn, Cd) and the hyperhoneycomb $\beta$-ZnIrO$_3$. The vertical dashed line indicates the XAS peak energy at $h\nu=11.217$ keV. The vertical purple line indicates $h\nu=11.214$ keV used for the RIXS measurements. (b) Ir $L_3$-edge resonant inelastic x-ray scattering (RIXS) spectra. The $J=3/2$ transitions are indicated by the vertical bars. The circles and triangles indicate the main peak and shoulder structures of the crystal field transitions to the $t_{2g}^{4}e_g$ multiplets.}
  \label{fig:rixs}
\end{figure}

\begin{figure*}[ht]
  \centering
  \includegraphics[width=\linewidth]{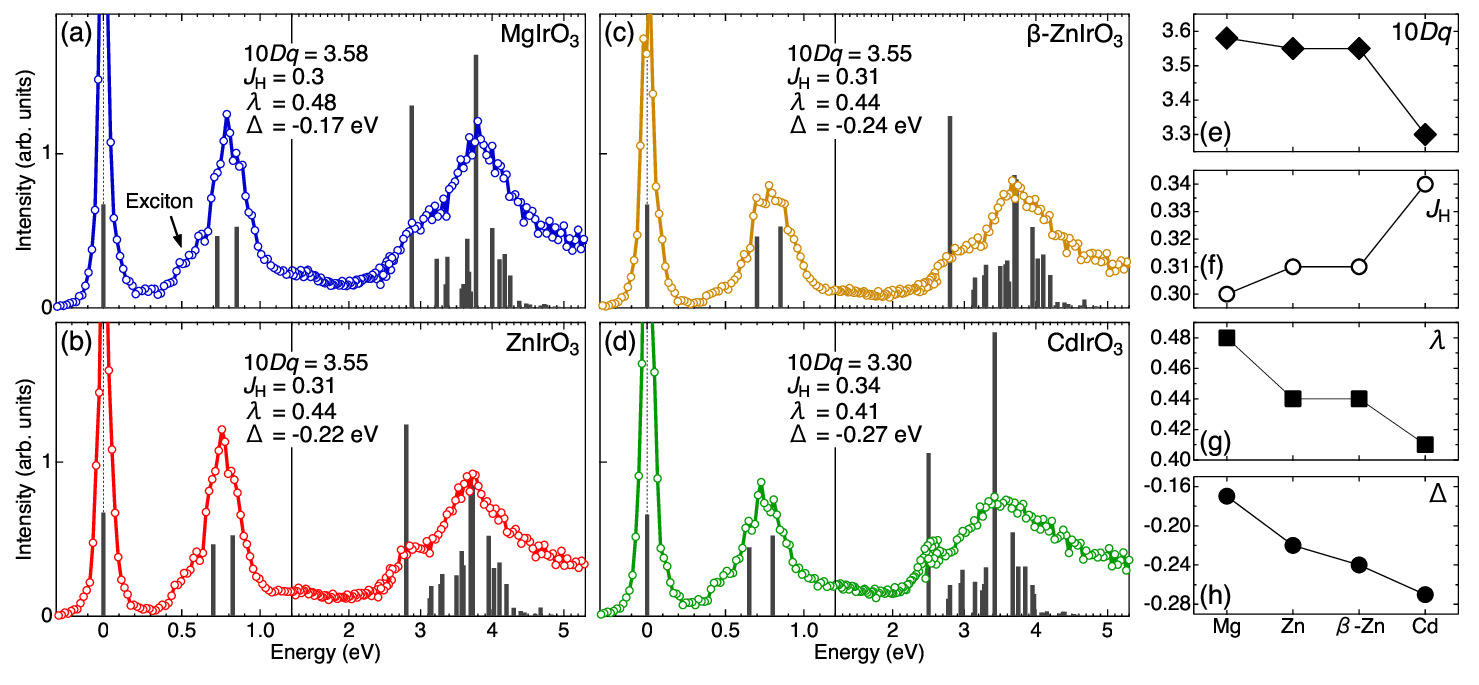}
  \caption{(a)-(d) Multiplet analysis of the RIXS spectra. Note that the spectra are divided into two regions below and above 1.2 eV for the better visualization of key spectral features. The ionic $d^5$ Hamiltonian [Eqs. (1)-(4)] includes the octahedral crystal field (10$Dq$), Hund's coupling ($J_H$), spin-orbit coupling ($\lambda$), and trigonal field ($\Delta$). The transition amplitudes for the optimal parameter sets are shown as gray vertical bars. (e)-(h) The evolution of the multiplet parameters.}
  \label{fig:multi}
\end{figure*}

\begin{align}
  &H_\mathrm{C} = U\sum_m n_{m\uparrow}n_{m\downarrow} + \sum_{m \neq m'} U'_{mm'} n_{m\uparrow}n_{m'\downarrow}\notag\\
   &\qquad\quad + \sum_{m<m'}\sum_{\sigma} (U'_{mm'} - J_{\mathrm{H},mm'}) n_{m\sigma}n_{m'\sigma}\notag\\
   &\qquad\quad - \sum_{m \neq m'} J_{\mathrm{H},mm'} d^{\dagger}_{m\uparrow}d_{m\downarrow}d^{\dagger}_{m'\downarrow}d_{m'\uparrow}\notag\\
   &\qquad\quad + \sum_{m \neq m'} J_{\mathrm{H},mm'} d^{\dagger}_{m\uparrow}d^{\dagger}_{m\downarrow}d_{m'\downarrow}d_{m'\uparrow}\\ \notag
   &\qquad\quad + \sum_{\mathrm{3\ or\ 4\ orb.}}\sum_{\sigma\sigma^\prime} U_{m_1m_2m_3m_4} d^{\dagger}_{m_1\sigma}d^{\dagger}_{m_2\sigma^\prime}d_{m_3\sigma^\prime}d_{m_4\sigma},\\
  &H_\mathrm{SOC} = \lambda \sum_i \bm{l}_i \cdot \bm{s}_i,\\ 
  &H_\mathrm{cub} = 10Dq\left[\frac{3}{5}n_{e_g} - \frac{2}{5}n_{t_{2g}} \right],\\
  &H_\mathrm{trig} = \Delta \left[ \frac{1}{3}n_{e_{g}^{\prime}}-\frac{2}{3}n_{a_{1g}} \right],
\end{align}
where $d^{\dagger}_{m\sigma}$ and $n_{m\sigma}$ are the creation operator for the $m$ orbital with spin $\sigma$ and the corresponding number operators, respectively. The last term in $H_\mathrm{C}$ consists of Coulomb interaction terms involving three or four different 5$d$ orbitals \cite{Sugano.S_etal.1970}. The trigonal field $\Delta$ splits the $t_{2g}$ orbitals into the $e_g^\prime$ doublet and $a_{1g}$ singlet, where the positive (negative) $\Delta$ corresponds to the elongation (compression) of the octahedra along the trigonal axis ($\parallel c$ axis for ilmenites) \cite{Chaloupka.J_etal.Phys.-Rev.-B2016}. We employ the spherical symmetry approximation of the interaction terms in $H_\mathrm{C}$, which imposes the constraint $U'_{mm'} = U - 2J_{\mathrm{H},mm'}$. The multiplet energy levels from the ground state then become independent of $U$ in the ionic model with the fixed electron number $d^5$. Hund's coupling parameters, $J_{\mathrm{H},mm'}$, are expressed in terms of Racah parameters. We have fixed the ratio $C/B = 5$ between the Racah parameters $B$ and $C$ \cite{Sugano.S_etal.1970}. Then four independent parameters remain: $10Dq$, $J_H$ between the $t_{2g}$ orbitals, $\lambda$, and $\Delta$. These parameters are optimized to reproduce the peak energies of the characteristic multiplet features in the RIXS spectra in Fig. \ref{fig:rixs}(b). Note that the determined parameters represent effective values that incorporate effects not explicitly included in the ionic model. For example, $J_H$ is strongly renormalized from the atomic value by the interaction with the charge continuum and the hybridization of the $5d$ orbitals with the ligand $p$ orbitals. Also, $\lambda$ can be reduced from the atomic value due to the covalency effect \cite{Abragam.A_etal.1970}. The RIXS transition amplitude was computed within the fast-collision approximation \cite{Ament.L_etal.Rev.-Mod.-Phys.2011}, and powder averaging of the RIXS intensity is performed by averaging the transition amplitudes over all diffraction angles, as outlined in Ref. \cite{Gretarsson.H_etal.Phys.-Rev.-B2024}.

Figures \ref{fig:multi}(a)-(d) compare the RIXS spectra and the calculated transition amplitudes for $A$IrO$_3$ and $\beta$-ZnIrO$_3$. Note that the spectra are presented in two separate panels, below and above 1.2 eV, to better visualize the lineshapes of the low-energy $J=3/2$ transition and the high-energy crystal-field transitions, respectively. The determined optimal parameter sets are indicated in the insets. The theoretical RIXS transition amplitudes with these optimal parameter sets are shown as vertical bars. 

Below 1.2 eV, one readily identifies the gradual decrease in the peak energy of the $J=3/2$ multiplet peak with increasing $A$-site ion size in $A$IrO$_3$. As the energy of this peak is located approximately at $3\lambda$/2, this results in the decreasing trend in the spin-orbit coupling parameter $\lambda$ [Fig. \ref{fig:multi}(g)]. Concomitantly, it shows a gradual broadening, suggesting the increasing crystal field distortion, leading to the gradual increase in the absolute value of the trigonal field $\Delta$ [Fig. \ref{fig:multi}(h)]. While the sign of $\Delta$ cannot be unambiguously determined from the angle-averaged measurement of powder samples, the compressed crystal structure of these iridates strongly suggests the negative sign of $\Delta$. Note that the shoulder structure identified around 0.5 eV [see panel (a)] is ascribed to the exciton bound states, which are also identified in $A_2$IrO$_3$ ($A=$ Na, Li) \cite{Gretarsson.H_etal.Phys.-Rev.-Lett.2013} and Ru$X_3$ ($X=$ Cl, Br) \cite{Lebert.B_etal.Phys.-Rev.-B2023}.

Above 1.2 eV, the crystal field transitions to the $t_{2g}^4e_g$ multiplets are observed. The main peak at $\sim$ 3.5 eV and shoulder structures at $\sim$ 2.8 eV are properly captured by the calculations. The gradual decrease in the peak energies of the main peak with increasing $A$-site ionic radii is ascribed to the decreasing octahedral crystal field parameter $10Dq$ [Fig. \ref{fig:multi}(e)]. On the other hand, the energy separation of these two peaks is primarily determined by the Hund's coupling parameter $J_H$. The gradual separation of the two features and the broadening of the main peak result in the increasing trend in $J_H$ [Fig. \ref{fig:multi}(f)].

The multiplet parameters obtained from the analysis are summarized in Fig. \ref{fig:multi}(e)-(h). The octahedral crystal field parameter $10Dq$ shows a gradual decrease with increasing $A$-site ionic radii, which is consistent with the increased Ir-O bond lengths. The Hund's coupling parameter $J_H$ shows a gradual increase, which may be ascribed to the increasing localization of the Ir $5d$ electrons. The spin-orbit coupling parameter $\lambda$ shows a decreasing trend. As the atomic SOC constant of the iridium ions is identical, this indicates the increased Ir $5d$ - O $2p$ covalency. This is consistent with the overall broadening of multiplet features [see Fig. \ref{fig:multi}(b)]. The trigonal field parameter $\Delta$ shows a gradual increase in its absolute value, which is consistent with the increasing distortion of the IrO$_6$ octahedra from the cubic symmetry.

Notably, the multiplet parameters for the ilmenite $\mathrm{ZnIrO}_3$ and the hyperhoneycomb $\beta\text{-}\mathrm{ZnIrO}_3$ are found to be nearly identical, confirming that their local crystal field environments and intraionic interactions are almost identical. Consequently, the contrast in their magnetic ground states originates from their different lattice structures, rather than from the difference in their local single-ion properties.

The systematic evolution of the multiplet parameters across the $A\mathrm{IrO}_3$ series provides a solid microscopic basis for understanding the modifications in the magnetic Hamiltonian. Specifically, the enhanced trigonal field absolute value $|\Delta|$ and the reduced spin-orbit coupling $\lambda$ with larger $A$-site ionic radii promote the mixing between the $J=1/2$ and $J=3/2$ states. This mixing naturally accounts for the significant deviation of the effective magnetic moment from the ideal $J=1/2$ value (e.g., $2.26~\mu_B$ in $\mathrm{CdIrO}_3$) and the consequent enhancement of non-Kitaev interactions. Thus, our RIXS observations microscopically validate the increasing antiferromagnetic transition temperatures in $A\mathrm{IrO}_3$ with increasing $A$-site size, reinforcing the notion that minimizing the local trigonal distortion is a key prerequisite for realizing the Kitaev spin liquid.

\section{Conclusion}


In conclusion, we have determined the multiplet structures of the ilmenite iridates $A\mathrm{IrO}_3$ ($A=\mathrm{Mg}, \mathrm{Zn}, \mathrm{Cd}$) and the hyperhoneycomb $\beta\text{-}\mathrm{ZnIrO}_3$ using Ir $L_3$-edge RIXS. By quantitatively extracting the relevant energy scales from the multiplet analysis, we have shown that the increasing $A$-site ionic radius from Mg to Cd leads to a decreasing $10Dq$ and a more pronounced trigonal-field splitting. This enhanced trigonal crystal field promotes the mixing of the $J=1/2$ and $J=3/2$ states and partially quenches the orbital angular momentum. These results suggest the enhanced non-Kitaev interactions and the deviation from the pure $J=1/2$ state in $\mathrm{CdIrO}_3$. Additionally, the nearly identical local electronic environments of the ilmenite $\mathrm{ZnIrO}_3$ and the hyperhoneycomb $\beta\text{-}\mathrm{ZnIrO}_3$ confirm that their distinct magnetic ground states stem from their different lattice structures rather than the local single-ion properties. Our results highlight the critical role of local crystal-field distortions in tuning the magnetic interactions in spin-orbit-coupled Mott insulators, offering guidelines for the design of Kitaev quantum spin liquids.

The authors thank H. Gretarsson for enlightening discussions. This work was supported by Grants-in-Aid for Scientific Research from JSPS (KAKENHI) (numbers JP22K13994, JP25K00014, JP25K01496), and JST PRESTO Grant Number JPMJPR23Q8. This work was supported by the QST Advanced Research Infrastructure for Materials and Nanotechnology of the Ministry of Education, Culture, Sports, Science and Technology (MEXT), Japan (Grant No. JPMXP1223QS0004). The synchrotron radiation experiments were performed using a QST experimental station at QST beamline BL11XU, SPring-8, with the approval of the Japan Synchrotron Radiation Research Institute (JASRI) (Proposal No.2023A3596).

\section*{Data Availability}
The data that support the findings of this article are openly available \cite{Zenodo}.

\bibliography{IlmeniteIr}

\begin{thebibliography}{35}%
\makeatletter
\providecommand \@ifxundefined [1]{%
 \@ifx{#1\undefined}
}%
\providecommand \@ifnum [1]{%
 \ifnum #1\expandafter \@firstoftwo
 \else \expandafter \@secondoftwo
 \fi
}%
\providecommand \@ifx [1]{%
 \ifx #1\expandafter \@firstoftwo
 \else \expandafter \@secondoftwo
 \fi
}%
\providecommand \natexlab [1]{#1}%
\providecommand \enquote  [1]{``#1''}%
\providecommand \bibnamefont  [1]{#1}%
\providecommand \bibfnamefont [1]{#1}%
\providecommand \citenamefont [1]{#1}%
\providecommand \href@noop [0]{\@secondoftwo}%
\providecommand \href [0]{\begingroup \@sanitize@url \@href}%
\providecommand \@href[1]{\@@startlink{#1}\@@href}%
\providecommand \@@href[1]{\endgroup#1\@@endlink}%
\providecommand \@sanitize@url [0]{\catcode `\\12\catcode `\$12\catcode
  `\&12\catcode `\#12\catcode `\^12\catcode `\_12\catcode `\%12\relax}%
\providecommand \@@startlink[1]{}%
\providecommand \@@endlink[0]{}%
\providecommand \url  [0]{\begingroup\@sanitize@url \@url }%
\providecommand \@url [1]{\endgroup\@href {#1}{\urlprefix }}%
\providecommand \urlprefix  [0]{URL }%
\providecommand \Eprint [0]{\href }%
\providecommand \doibase [0]{https://doi.org/}%
\providecommand \selectlanguage [0]{\@gobble}%
\providecommand \bibinfo  [0]{\@secondoftwo}%
\providecommand \bibfield  [0]{\@secondoftwo}%
\providecommand \translation [1]{[#1]}%
\providecommand \BibitemOpen [0]{}%
\providecommand \bibitemStop [0]{}%
\providecommand \bibitemNoStop [0]{.\EOS\space}%
\providecommand \EOS [0]{\spacefactor3000\relax}%
\providecommand \BibitemShut  [1]{\csname bibitem#1\endcsname}%
\let\auto@bib@innerbib\@empty
\bibitem [{\citenamefont {Kitaev}(2006)}]{Kitaev.A_etal.Ann.-Phys.2006}%
  \BibitemOpen
  \bibfield  {author} {\bibinfo {author} {\bibfnamefont {A.}~\bibnamefont
  {Kitaev}},\ }\bibfield  {title} {\bibinfo {title} {Anyons in an exactly
  solved model and beyond},\ }\href
  {https://doi.org/https://doi.org/10.1016/j.aop.2005.10.005} {\bibfield
  {journal} {\bibinfo  {journal} {Ann. Phys.}\ }\textbf {\bibinfo {volume}
  {321}},\ \bibinfo {pages} {2 } (\bibinfo {year} {2006})}\BibitemShut
  {NoStop}%
\bibitem [{\citenamefont {Savary}\ and\ \citenamefont
  {Balents}(2017)}]{Savary.L_etal.Rep.-Prog.-Phys.2017}%
  \BibitemOpen
  \bibfield  {author} {\bibinfo {author} {\bibfnamefont {L.}~\bibnamefont
  {Savary}}\ and\ \bibinfo {author} {\bibfnamefont {L.}~\bibnamefont
  {Balents}},\ }\bibfield  {title} {\bibinfo {title} {Quantum spin liquids: a
  review},\ }\href {http://stacks.iop.org/0034-4885/80/i=1/a=016502} {\bibfield
   {journal} {\bibinfo  {journal} {Rep. Prog. Phys.}\ }\textbf {\bibinfo
  {volume} {80}},\ \bibinfo {pages} {016502} (\bibinfo {year}
  {2017})}\BibitemShut {NoStop}%
\bibitem [{\citenamefont {Jackeli}\ and\ \citenamefont
  {Khaliullin}(2009)}]{Jackeli.G_etal.Phys.-Rev.-Lett.2009}%
  \BibitemOpen
  \bibfield  {author} {\bibinfo {author} {\bibfnamefont {G.}~\bibnamefont
  {Jackeli}}\ and\ \bibinfo {author} {\bibfnamefont {G.}~\bibnamefont
  {Khaliullin}},\ }\bibfield  {title} {\bibinfo {title} {Mott insulators in the
  strong spin-orbit coupling limit: From Heisenberg to a quantum compass and
  Kitaev models},\ }\href {https://doi.org/10.1103/PhysRevLett.102.017205}
  {\bibfield  {journal} {\bibinfo  {journal} {Phys. Rev. Lett.}\ }\textbf
  {\bibinfo {volume} {102}},\ \bibinfo {pages} {017205} (\bibinfo {year}
  {2009})}\BibitemShut {NoStop}%
\bibitem [{\citenamefont {Rau}\ \emph {et~al.}(2016)\citenamefont {Rau},
  \citenamefont {Lee},\ and\ \citenamefont
  {Kee}}]{Rau.J_etal.Annu.-Rev.-Condens.-Matter-Phys.2016}%
  \BibitemOpen
  \bibfield  {author} {\bibinfo {author} {\bibfnamefont {J.~G.}\ \bibnamefont
  {Rau}}, \bibinfo {author} {\bibfnamefont {E.~K.-H.}\ \bibnamefont {Lee}},\
  and\ \bibinfo {author} {\bibfnamefont {H.-Y.}\ \bibnamefont {Kee}},\
  }\bibfield  {title} {\bibinfo {title} {Spin-orbit physics giving rise to
  novel phases in correlated systems: Iridates and related materials},\ }\href
  {https://doi.org/10.1146/annurev-conmatphys-031115-011319} {\bibfield
  {journal} {\bibinfo  {journal} {Annu. Rev. Condens. Matter Phys.}\ }\textbf
  {\bibinfo {volume} {7}},\ \bibinfo {pages} {195} (\bibinfo {year}
  {2016})}\BibitemShut {NoStop}%
\bibitem [{\citenamefont {Hermanns}\ \emph {et~al.}(2018)\citenamefont
  {Hermanns}, \citenamefont {Kimchi},\ and\ \citenamefont
  {Knolle}}]{Hermanns.M_etal.Annu.-Rev.-Condens.-Matter-Phys.2018}%
  \BibitemOpen
  \bibfield  {author} {\bibinfo {author} {\bibfnamefont {M.}~\bibnamefont
  {Hermanns}}, \bibinfo {author} {\bibfnamefont {I.}~\bibnamefont {Kimchi}},\
  and\ \bibinfo {author} {\bibfnamefont {J.}~\bibnamefont {Knolle}},\
  }\bibfield  {title} {\bibinfo {title} {Physics of the Kitaev model:
  Fractionalization, dynamic correlations, and material connections},\ }\href
  {https://doi.org/10.1146/annurev-conmatphys-033117-053934} {\bibfield
  {journal} {\bibinfo  {journal} {Annu. Rev. Condens. Matter Phys.}\ }\textbf
  {\bibinfo {volume} {9}},\ \bibinfo {pages} {17} (\bibinfo {year}
  {2018})}\BibitemShut {NoStop}%
\bibitem [{\citenamefont {Takagi}\ \emph {et~al.}(2019)\citenamefont {Takagi},
  \citenamefont {Takayama}, \citenamefont {Jackeli}, \citenamefont
  {Khaliullin},\ and\ \citenamefont
  {Nagler}}]{Takagi.H_etal.Nat.-Rev.-Phys.2019}%
  \BibitemOpen
  \bibfield  {author} {\bibinfo {author} {\bibfnamefont {H.}~\bibnamefont
  {Takagi}}, \bibinfo {author} {\bibfnamefont {T.}~\bibnamefont {Takayama}},
  \bibinfo {author} {\bibfnamefont {G.}~\bibnamefont {Jackeli}}, \bibinfo
  {author} {\bibfnamefont {G.}~\bibnamefont {Khaliullin}},\ and\ \bibinfo
  {author} {\bibfnamefont {S.~E.}\ \bibnamefont {Nagler}},\ }\bibfield  {title}
  {\bibinfo {title} {Concept and realization of Kitaev quantum spin liquids},\
  }\href {https://doi.org/10.1038/s42254-019-0038-2} {\bibfield  {journal}
  {\bibinfo  {journal} {Nat. Rev. Phys.}\ }\textbf {\bibinfo {volume} {1}},\
  \bibinfo {pages} {264} (\bibinfo {year} {2019})}\BibitemShut {NoStop}%
\bibitem [{\citenamefont {Motome}\ \emph {et~al.}(2020)\citenamefont {Motome},
  \citenamefont {Sano}, \citenamefont {Jang}, \citenamefont {Sugita},\ and\
  \citenamefont {Kato}}]{Motome.Y_etal.J.-Phys.-Condens.-Matter2020}%
  \BibitemOpen
  \bibfield  {author} {\bibinfo {author} {\bibfnamefont {Y.}~\bibnamefont
  {Motome}}, \bibinfo {author} {\bibfnamefont {R.}~\bibnamefont {Sano}},
  \bibinfo {author} {\bibfnamefont {S.}~\bibnamefont {Jang}}, \bibinfo {author}
  {\bibfnamefont {Y.}~\bibnamefont {Sugita}},\ and\ \bibinfo {author}
  {\bibfnamefont {Y.}~\bibnamefont {Kato}},\ }\bibfield  {title} {\bibinfo
  {title} {Materials design of Kitaev spin liquids beyond the
  Jackeli{\textendash}Khaliullin mechanism},\ }\href
  {https://doi.org/10.1088/1361-648x/ab8525} {\bibfield  {journal} {\bibinfo
  {journal} {J. Phys. Condens. Matter}\ }\textbf {\bibinfo {volume} {32}},\
  \bibinfo {pages} {404001} (\bibinfo {year} {2020})}\BibitemShut {NoStop}%
\bibitem [{\citenamefont {Haraguchi}\ \emph {et~al.}(2018)\citenamefont
  {Haraguchi}, \citenamefont {Michioka}, \citenamefont {Matsuo}, \citenamefont
  {Kindo}, \citenamefont {Ueda},\ and\ \citenamefont
  {Yoshimura}}]{Haraguchi.Y_etal.Phys.-Rev.-Materials2018}%
  \BibitemOpen
  \bibfield  {author} {\bibinfo {author} {\bibfnamefont {Y.}~\bibnamefont
  {Haraguchi}}, \bibinfo {author} {\bibfnamefont {C.}~\bibnamefont {Michioka}},
  \bibinfo {author} {\bibfnamefont {A.}~\bibnamefont {Matsuo}}, \bibinfo
  {author} {\bibfnamefont {K.}~\bibnamefont {Kindo}}, \bibinfo {author}
  {\bibfnamefont {H.}~\bibnamefont {Ueda}},\ and\ \bibinfo {author}
  {\bibfnamefont {K.}~\bibnamefont {Yoshimura}},\ }\bibfield  {title} {\bibinfo
  {title} {Magnetic ordering with an XY-like anisotropy in the honeycomb
  lattice iridates ${\mathrm{ZnIrO}}_{3}$ and ${\mathrm{MgIrO}}_{3}$
  synthesized via a metathesis reaction},\ }\href
  {https://doi.org/10.1103/PhysRevMaterials.2.054411} {\bibfield  {journal}
  {\bibinfo  {journal} {Phys. Rev. Materials}\ }\textbf {\bibinfo {volume}
  {2}},\ \bibinfo {pages} {054411} (\bibinfo {year} {2018})}\BibitemShut
  {NoStop}%
\bibitem [{\citenamefont {Haraguchi}\ and\ \citenamefont
  {Katori}(2020)}]{Haraguchi.Y_etal.Phys.-Rev.-Materials2020}%
  \BibitemOpen
  \bibfield  {author} {\bibinfo {author} {\bibfnamefont {Y.}~\bibnamefont
  {Haraguchi}}\ and\ \bibinfo {author} {\bibfnamefont {H.~A.}\ \bibnamefont
  {Katori}},\ }\bibfield  {title} {\bibinfo {title} {Strong antiferromagnetic
  interaction owing to a large trigonal distortion in the spin-orbit-coupled
  honeycomb lattice iridate $\mathrm{CdIr}{\mathrm{O}}_{3}$},\ }\href
  {https://doi.org/10.1103/PhysRevMaterials.4.044401} {\bibfield  {journal}
  {\bibinfo  {journal} {Phys. Rev. Materials}\ }\textbf {\bibinfo {volume}
  {4}},\ \bibinfo {pages} {044401} (\bibinfo {year} {2020})}\BibitemShut
  {NoStop}%
\bibitem [{\citenamefont {Singh}\ and\ \citenamefont
  {Gegenwart}(2010)}]{Singh.Y_etal.Phys.-Rev.-B2010}%
  \BibitemOpen
  \bibfield  {author} {\bibinfo {author} {\bibfnamefont {Y.}~\bibnamefont
  {Singh}}\ and\ \bibinfo {author} {\bibfnamefont {P.}~\bibnamefont
  {Gegenwart}},\ }\bibfield  {title} {\bibinfo {title} {Antiferromagnetic Mott
  insulating state in single crystals of the honeycomb lattice material
  ${\text{Na}}_{2}{\text{IrO}}_{3}$},\ }\href
  {https://doi.org/10.1103/PhysRevB.82.064412} {\bibfield  {journal} {\bibinfo
  {journal} {Phys. Rev. B}\ }\textbf {\bibinfo {volume} {82}},\ \bibinfo
  {pages} {064412} (\bibinfo {year} {2010})}\BibitemShut {NoStop}%
\bibitem [{\citenamefont {Hwan~Chun}\ \emph {et~al.}(2015)\citenamefont
  {Hwan~Chun}, \citenamefont {Kim}, \citenamefont {Kim}, \citenamefont {Zheng},
  \citenamefont {Stoumpos}, \citenamefont {Malliakas}, \citenamefont
  {Mitchell}, \citenamefont {Mehlawat}, \citenamefont {Singh}, \citenamefont
  {Choi}, \citenamefont {Gog}, \citenamefont {Al-Zein}, \citenamefont {Sala},
  \citenamefont {Krisch}, \citenamefont {Chaloupka}, \citenamefont {Jackeli},
  \citenamefont {Khaliullin},\ and\ \citenamefont
  {Kim}}]{Hwan-Chun.S_etal.Nat.-Phys.2015}%
  \BibitemOpen
  \bibfield  {author} {\bibinfo {author} {\bibfnamefont {S.}~\bibnamefont
  {Hwan~Chun}}, \bibinfo {author} {\bibfnamefont {J.-W.}\ \bibnamefont {Kim}},
  \bibinfo {author} {\bibfnamefont {J.}~\bibnamefont {Kim}}, \bibinfo {author}
  {\bibfnamefont {H.}~\bibnamefont {Zheng}}, \bibinfo {author} {\bibfnamefont
  {C.~C.}\ \bibnamefont {Stoumpos}}, \bibinfo {author} {\bibfnamefont {C.~D.}\
  \bibnamefont {Malliakas}}, \bibinfo {author} {\bibfnamefont {J.~F.}\
  \bibnamefont {Mitchell}}, \bibinfo {author} {\bibfnamefont {K.}~\bibnamefont
  {Mehlawat}}, \bibinfo {author} {\bibfnamefont {Y.}~\bibnamefont {Singh}},
  \bibinfo {author} {\bibfnamefont {Y.}~\bibnamefont {Choi}}, \bibinfo {author}
  {\bibfnamefont {T.}~\bibnamefont {Gog}}, \bibinfo {author} {\bibfnamefont
  {A.}~\bibnamefont {Al-Zein}}, \bibinfo {author} {\bibfnamefont {M.~M.}\
  \bibnamefont {Sala}}, \bibinfo {author} {\bibfnamefont {M.}~\bibnamefont
  {Krisch}}, \bibinfo {author} {\bibfnamefont {J.}~\bibnamefont {Chaloupka}},
  \bibinfo {author} {\bibfnamefont {G.}~\bibnamefont {Jackeli}}, \bibinfo
  {author} {\bibfnamefont {G.}~\bibnamefont {Khaliullin}},\ and\ \bibinfo
  {author} {\bibfnamefont {B.~J.}\ \bibnamefont {Kim}},\ }\bibfield  {title}
  {\bibinfo {title} {Direct evidence for dominant bond-directional interactions
  in a honeycomb lattice iridate Na$_2$IrO$_3$},\ }\href
  {http://dx.doi.org/10.1038/nphys3322} {\bibfield  {journal} {\bibinfo
  {journal} {Nat. Phys.}\ }\textbf {\bibinfo {volume} {11}},\ \bibinfo {pages}
  {462 EP } (\bibinfo {year} {2015})}\BibitemShut {NoStop}%
\bibitem [{\citenamefont {Singh}\ \emph {et~al.}(2012)\citenamefont {Singh},
  \citenamefont {Manni}, \citenamefont {Reuther}, \citenamefont {Berlijn},
  \citenamefont {Thomale}, \citenamefont {Ku}, \citenamefont {Trebst},\ and\
  \citenamefont {Gegenwart}}]{Singh.Y_etal.Phys.-Rev.-Lett.2012}%
  \BibitemOpen
  \bibfield  {author} {\bibinfo {author} {\bibfnamefont {Y.}~\bibnamefont
  {Singh}}, \bibinfo {author} {\bibfnamefont {S.}~\bibnamefont {Manni}},
  \bibinfo {author} {\bibfnamefont {J.}~\bibnamefont {Reuther}}, \bibinfo
  {author} {\bibfnamefont {T.}~\bibnamefont {Berlijn}}, \bibinfo {author}
  {\bibfnamefont {R.}~\bibnamefont {Thomale}}, \bibinfo {author} {\bibfnamefont
  {W.}~\bibnamefont {Ku}}, \bibinfo {author} {\bibfnamefont {S.}~\bibnamefont
  {Trebst}},\ and\ \bibinfo {author} {\bibfnamefont {P.}~\bibnamefont
  {Gegenwart}},\ }\bibfield  {title} {\bibinfo {title} {Relevance of the
  Heisenberg-Kitaev model for the honeycomb lattice iridates
  ${A}_{2}{\mathrm{IrO}}_{3}$},\ }\href
  {https://doi.org/10.1103/PhysRevLett.108.127203} {\bibfield  {journal}
  {\bibinfo  {journal} {Phys. Rev. Lett.}\ }\textbf {\bibinfo {volume} {108}},\
  \bibinfo {pages} {127203} (\bibinfo {year} {2012})}\BibitemShut {NoStop}%
\bibitem [{\citenamefont {Choi}\ \emph {et~al.}(2019)\citenamefont {Choi},
  \citenamefont {Lee}, \citenamefont {Lee}, \citenamefont {Yoon}, \citenamefont
  {Lee}, \citenamefont {Park}, \citenamefont {Ali}, \citenamefont {Singh},
  \citenamefont {Orain}, \citenamefont {Kim}, \citenamefont {Rhyee},
  \citenamefont {Chen}, \citenamefont {Chou},\ and\ \citenamefont
  {Choi}}]{Choi.Y_etal.Phys.-Rev.-Lett.2019}%
  \BibitemOpen
  \bibfield  {author} {\bibinfo {author} {\bibfnamefont {Y.~S.}\ \bibnamefont
  {Choi}}, \bibinfo {author} {\bibfnamefont {C.~H.}\ \bibnamefont {Lee}},
  \bibinfo {author} {\bibfnamefont {S.}~\bibnamefont {Lee}}, \bibinfo {author}
  {\bibfnamefont {S.}~\bibnamefont {Yoon}}, \bibinfo {author} {\bibfnamefont
  {W.-J.}\ \bibnamefont {Lee}}, \bibinfo {author} {\bibfnamefont
  {J.}~\bibnamefont {Park}}, \bibinfo {author} {\bibfnamefont {A.}~\bibnamefont
  {Ali}}, \bibinfo {author} {\bibfnamefont {Y.}~\bibnamefont {Singh}}, \bibinfo
  {author} {\bibfnamefont {J.-C.}\ \bibnamefont {Orain}}, \bibinfo {author}
  {\bibfnamefont {G.}~\bibnamefont {Kim}}, \bibinfo {author} {\bibfnamefont
  {J.-S.}\ \bibnamefont {Rhyee}}, \bibinfo {author} {\bibfnamefont {W.-T.}\
  \bibnamefont {Chen}}, \bibinfo {author} {\bibfnamefont {F.}~\bibnamefont
  {Chou}},\ and\ \bibinfo {author} {\bibfnamefont {K.-Y.}\ \bibnamefont
  {Choi}},\ }\bibfield  {title} {\bibinfo {title} {Exotic low-energy
  excitations emergent in the random Kitaev magnet
  ${\mathrm{Cu}}_{2}{\mathrm{IrO}}_{3}$},\ }\href
  {https://doi.org/10.1103/PhysRevLett.122.167202} {\bibfield  {journal}
  {\bibinfo  {journal} {Phys. Rev. Lett.}\ }\textbf {\bibinfo {volume} {122}},\
  \bibinfo {pages} {167202} (\bibinfo {year} {2019})}\BibitemShut {NoStop}%
\bibitem [{\citenamefont {Haraguchi}\ \emph {et~al.}(2024)\citenamefont
  {Haraguchi}, \citenamefont {Nishio-Hamane}, \citenamefont {Matsuo},
  \citenamefont {Kindo},\ and\ \citenamefont
  {Katori}}]{Haraguchi.Y_etal.J.-Phys.-Condens.-Matter2024}%
  \BibitemOpen
  \bibfield  {author} {\bibinfo {author} {\bibfnamefont {Y.}~\bibnamefont
  {Haraguchi}}, \bibinfo {author} {\bibfnamefont {D.}~\bibnamefont
  {Nishio-Hamane}}, \bibinfo {author} {\bibfnamefont {A.}~\bibnamefont
  {Matsuo}}, \bibinfo {author} {\bibfnamefont {K.}~\bibnamefont {Kindo}},\ and\
  \bibinfo {author} {\bibfnamefont {H.~A.}\ \bibnamefont {Katori}},\ }\bibfield
   {title} {\bibinfo {title} {High-temperature magnetic anomaly via suppression
  of antisite disorder through synthesis route modification in a Kitaev
  candidate Cu$_2$IrO$_3$},\ }\href {https://doi.org/10.1088/1361-648X/ad5d3a}
  {\bibfield  {journal} {\bibinfo  {journal} {J. Phys. Condens. Matter}\
  }\textbf {\bibinfo {volume} {36}},\ \bibinfo {pages} {405801} (\bibinfo
  {year} {2024})}\BibitemShut {NoStop}%
\bibitem [{\citenamefont {Kitagawa}\ \emph {et~al.}(2018)\citenamefont
  {Kitagawa}, \citenamefont {Takayama}, \citenamefont {Matsumoto},
  \citenamefont {Kato}, \citenamefont {Takano}, \citenamefont {Kishimoto},
  \citenamefont {Bette}, \citenamefont {Dinnebier}, \citenamefont {Jackeli},\
  and\ \citenamefont {Takagi}}]{Kitagawa.K_etal.Nature2018}%
  \BibitemOpen
  \bibfield  {author} {\bibinfo {author} {\bibfnamefont {K.}~\bibnamefont
  {Kitagawa}}, \bibinfo {author} {\bibfnamefont {T.}~\bibnamefont {Takayama}},
  \bibinfo {author} {\bibfnamefont {Y.}~\bibnamefont {Matsumoto}}, \bibinfo
  {author} {\bibfnamefont {A.}~\bibnamefont {Kato}}, \bibinfo {author}
  {\bibfnamefont {R.}~\bibnamefont {Takano}}, \bibinfo {author} {\bibfnamefont
  {Y.}~\bibnamefont {Kishimoto}}, \bibinfo {author} {\bibfnamefont
  {S.}~\bibnamefont {Bette}}, \bibinfo {author} {\bibfnamefont
  {R.}~\bibnamefont {Dinnebier}}, \bibinfo {author} {\bibfnamefont
  {G.}~\bibnamefont {Jackeli}},\ and\ \bibinfo {author} {\bibfnamefont
  {H.}~\bibnamefont {Takagi}},\ }\bibfield  {title} {\bibinfo {title} {A
  spin--orbital-entangled quantum liquid on a honeycomb lattice},\ }\href
  {https://doi.org/10.1038/nature25482} {\bibfield  {journal} {\bibinfo
  {journal} {Nature}\ }\textbf {\bibinfo {volume} {554}},\ \bibinfo {pages}
  {341} (\bibinfo {year} {2018})}\BibitemShut {NoStop}%
\bibitem [{\citenamefont {Bahrami}\ \emph {et~al.}(2019)\citenamefont
  {Bahrami}, \citenamefont {Lafargue-Dit-Hauret}, \citenamefont {Lebedev},
  \citenamefont {Movshovich}, \citenamefont {Yang}, \citenamefont {Broido},
  \citenamefont {Rocquefelte},\ and\ \citenamefont
  {Tafti}}]{Bahrami.F_etal.Phys.-Rev.-Lett.2019}%
  \BibitemOpen
  \bibfield  {author} {\bibinfo {author} {\bibfnamefont {F.}~\bibnamefont
  {Bahrami}}, \bibinfo {author} {\bibfnamefont {W.}~\bibnamefont
  {Lafargue-Dit-Hauret}}, \bibinfo {author} {\bibfnamefont {O.~I.}\
  \bibnamefont {Lebedev}}, \bibinfo {author} {\bibfnamefont {R.}~\bibnamefont
  {Movshovich}}, \bibinfo {author} {\bibfnamefont {H.-Y.}\ \bibnamefont
  {Yang}}, \bibinfo {author} {\bibfnamefont {D.}~\bibnamefont {Broido}},
  \bibinfo {author} {\bibfnamefont {X.}~\bibnamefont {Rocquefelte}},\ and\
  \bibinfo {author} {\bibfnamefont {F.}~\bibnamefont {Tafti}},\ }\bibfield
  {title} {\bibinfo {title} {Thermodynamic evidence of proximity to a Kitaev
  spin liquid in ${\mathrm{Ag}}_{3}{\mathrm{LiIr}}_{2}{\mathrm{O}}_{6}$},\
  }\href {https://doi.org/10.1103/PhysRevLett.123.237203} {\bibfield  {journal}
  {\bibinfo  {journal} {Phys. Rev. Lett.}\ }\textbf {\bibinfo {volume} {123}},\
  \bibinfo {pages} {237203} (\bibinfo {year} {2019})}\BibitemShut {NoStop}%
\bibitem [{\citenamefont {Plumb}\ \emph {et~al.}(2014)\citenamefont {Plumb},
  \citenamefont {Clancy}, \citenamefont {Sandilands}, \citenamefont {Shankar},
  \citenamefont {Hu}, \citenamefont {Burch}, \citenamefont {Kee},\ and\
  \citenamefont {Kim}}]{Plumb.K_etal.Phys.-Rev.-B2014}%
  \BibitemOpen
  \bibfield  {author} {\bibinfo {author} {\bibfnamefont {K.~W.}\ \bibnamefont
  {Plumb}}, \bibinfo {author} {\bibfnamefont {J.~P.}\ \bibnamefont {Clancy}},
  \bibinfo {author} {\bibfnamefont {L.~J.}\ \bibnamefont {Sandilands}},
  \bibinfo {author} {\bibfnamefont {V.~V.}\ \bibnamefont {Shankar}}, \bibinfo
  {author} {\bibfnamefont {Y.~F.}\ \bibnamefont {Hu}}, \bibinfo {author}
  {\bibfnamefont {K.~S.}\ \bibnamefont {Burch}}, \bibinfo {author}
  {\bibfnamefont {H.-Y.}\ \bibnamefont {Kee}},\ and\ \bibinfo {author}
  {\bibfnamefont {Y.-J.}\ \bibnamefont {Kim}},\ }\bibfield  {title} {\bibinfo
  {title} {$\ensuremath{\alpha}-{\mathrm{RuCl}}_{3}$: A spin-orbit assisted
  Mott insulator on a honeycomb lattice},\ }\href
  {https://doi.org/10.1103/PhysRevB.90.041112} {\bibfield  {journal} {\bibinfo
  {journal} {Phys. Rev. B}\ }\textbf {\bibinfo {volume} {90}},\ \bibinfo
  {pages} {041112} (\bibinfo {year} {2014})}\BibitemShut {NoStop}%
\bibitem [{\citenamefont {Sears}\ \emph {et~al.}(2015)\citenamefont {Sears},
  \citenamefont {Songvilay}, \citenamefont {Plumb}, \citenamefont {Clancy},
  \citenamefont {Qiu}, \citenamefont {Zhao}, \citenamefont {Parshall},\ and\
  \citenamefont {Kim}}]{Sears.J_etal.Phys.-Rev.-B2015}%
  \BibitemOpen
  \bibfield  {author} {\bibinfo {author} {\bibfnamefont {J.~A.}\ \bibnamefont
  {Sears}}, \bibinfo {author} {\bibfnamefont {M.}~\bibnamefont {Songvilay}},
  \bibinfo {author} {\bibfnamefont {K.~W.}\ \bibnamefont {Plumb}}, \bibinfo
  {author} {\bibfnamefont {J.~P.}\ \bibnamefont {Clancy}}, \bibinfo {author}
  {\bibfnamefont {Y.}~\bibnamefont {Qiu}}, \bibinfo {author} {\bibfnamefont
  {Y.}~\bibnamefont {Zhao}}, \bibinfo {author} {\bibfnamefont {D.}~\bibnamefont
  {Parshall}},\ and\ \bibinfo {author} {\bibfnamefont {Y.-J.}\ \bibnamefont
  {Kim}},\ }\bibfield  {title} {\bibinfo {title} {Magnetic order in
  $\ensuremath{\alpha}-{\text{RuCl}}_{3}$: A honeycomb-lattice quantum magnet
  with strong spin-orbit coupling},\ }\href
  {https://doi.org/10.1103/PhysRevB.91.144420} {\bibfield  {journal} {\bibinfo
  {journal} {Phys. Rev. B}\ }\textbf {\bibinfo {volume} {91}},\ \bibinfo
  {pages} {144420} (\bibinfo {year} {2015})}\BibitemShut {NoStop}%
\bibitem [{\citenamefont {Jang}\ and\ \citenamefont
  {Motome}(2021)}]{Jang.S_etal.Phys.-Rev.-Materials2021}%
  \BibitemOpen
  \bibfield  {author} {\bibinfo {author} {\bibfnamefont {S.-H.}\ \bibnamefont
  {Jang}}\ and\ \bibinfo {author} {\bibfnamefont {Y.}~\bibnamefont {Motome}},\
  }\bibfield  {title} {\bibinfo {title} {Electronic and magnetic properties of
  iridium ilmenites $A{\mathrm{IrO}}_{3}$ $(A=\mathrm{Mg}, \mathrm{Zn},
  \mathrm{and}$  $\mathrm{Mn})$},\ }\href
  {https://doi.org/10.1103/PhysRevMaterials.5.104409} {\bibfield  {journal}
  {\bibinfo  {journal} {Phys. Rev. Materials}\ }\textbf {\bibinfo {volume}
  {5}},\ \bibinfo {pages} {104409} (\bibinfo {year} {2021})}\BibitemShut
  {NoStop}%
\bibitem [{\citenamefont {Momma}\ and\ \citenamefont
  {Izumi}(2011)}]{Momma.K_etal.J.-Appl.-Cryst.2011}%
  \BibitemOpen
  \bibfield  {author} {\bibinfo {author} {\bibfnamefont {K.}~\bibnamefont
  {Momma}}\ and\ \bibinfo {author} {\bibfnamefont {F.}~\bibnamefont {Izumi}},\
  }\bibfield  {title} {\bibinfo {title} {{{\it VESTA3} for three-dimensional
  visualization of crystal, volumetric and morphology data}},\ }\href
  {https://doi.org/10.1107/S0021889811038970} {\bibfield  {journal} {\bibinfo
  {journal} {J. Appl. Cryst.}\ }\textbf {\bibinfo {volume} {44}},\ \bibinfo
  {pages} {1272} (\bibinfo {year} {2011})}\BibitemShut {NoStop}%
\bibitem [{\citenamefont {Haraguchi}\ \emph {et~al.}(2022)\citenamefont
  {Haraguchi}, \citenamefont {Matsuo}, \citenamefont {Kindo},\ and\
  \citenamefont {Katori}}]{Haraguchi.Y_etal.Phys.-Rev.-Materials2022}%
  \BibitemOpen
  \bibfield  {author} {\bibinfo {author} {\bibfnamefont {Y.}~\bibnamefont
  {Haraguchi}}, \bibinfo {author} {\bibfnamefont {A.}~\bibnamefont {Matsuo}},
  \bibinfo {author} {\bibfnamefont {K.}~\bibnamefont {Kindo}},\ and\ \bibinfo
  {author} {\bibfnamefont {H.~A.}\ \bibnamefont {Katori}},\ }\bibfield  {title}
  {\bibinfo {title} {Quantum paramagnetism in the hyperhoneycomb Kitaev magnet
  $\ensuremath{\beta}\text{\ensuremath{-}}{\mathrm{ZnIrO}}_{3}$},\ }\href
  {https://doi.org/10.1103/PhysRevMaterials.6.L021401} {\bibfield  {journal}
  {\bibinfo  {journal} {Phys. Rev. Materials}\ }\textbf {\bibinfo {volume}
  {6}},\ \bibinfo {pages} {L021401} (\bibinfo {year} {2022})}\BibitemShut
  {NoStop}%
\bibitem [{\citenamefont {Haraguchi}\ and\ \citenamefont
  {Katori}(2023)}]{Haraguchi.Y_etal.Chem.-Lett.2023}%
  \BibitemOpen
  \bibfield  {author} {\bibinfo {author} {\bibfnamefont {Y.}~\bibnamefont
  {Haraguchi}}\ and\ \bibinfo {author} {\bibfnamefont {H.~A.}\ \bibnamefont
  {Katori}},\ }\bibfield  {title} {\bibinfo {title} {Monoclinic distortion in
  hyperhoneycomb Kitaev material $\beta$-ZnIrO$_3$ revealed by improved sample
  quality},\ }\href {https://doi.org/10.1246/cl.230108} {\bibfield  {journal}
  {\bibinfo  {journal} {Chem. Lett.}\ }\textbf {\bibinfo {volume} {52}},\
  \bibinfo {pages} {404} (\bibinfo {year} {2023})}\BibitemShut {NoStop}%
\bibitem [{\citenamefont {Ament}\ \emph {et~al.}(2011)\citenamefont {Ament},
  \citenamefont {van Veenendaal}, \citenamefont {Devereaux}, \citenamefont
  {Hill},\ and\ \citenamefont {van~den
  Brink}}]{Ament.L_etal.Rev.-Mod.-Phys.2011}%
  \BibitemOpen
  \bibfield  {author} {\bibinfo {author} {\bibfnamefont {L.~J.~P.}\
  \bibnamefont {Ament}}, \bibinfo {author} {\bibfnamefont {M.}~\bibnamefont
  {van Veenendaal}}, \bibinfo {author} {\bibfnamefont {T.~P.}\ \bibnamefont
  {Devereaux}}, \bibinfo {author} {\bibfnamefont {J.~P.}\ \bibnamefont
  {Hill}},\ and\ \bibinfo {author} {\bibfnamefont {J.}~\bibnamefont {van~den
  Brink}},\ }\bibfield  {title} {\bibinfo {title} {Resonant inelastic x-ray
  scattering studies of elementary excitations},\ }\href
  {https://doi.org/10.1103/RevModPhys.83.705} {\bibfield  {journal} {\bibinfo
  {journal} {Rev. Mod. Phys.}\ }\textbf {\bibinfo {volume} {83}},\ \bibinfo
  {pages} {705} (\bibinfo {year} {2011})}\BibitemShut {NoStop}%
\bibitem [{\citenamefont {Ishii}\ \emph
  {et~al.}(2013{\natexlab{a}})\citenamefont {Ishii}, \citenamefont {Tohyama},\
  and\ \citenamefont {Mizuki}}]{Ishii.K_etal.J.-Phys.-Soc.-Jpn.2013}%
  \BibitemOpen
  \bibfield  {author} {\bibinfo {author} {\bibfnamefont {K.}~\bibnamefont
  {Ishii}}, \bibinfo {author} {\bibfnamefont {T.}~\bibnamefont {Tohyama}},\
  and\ \bibinfo {author} {\bibfnamefont {J.}~\bibnamefont {Mizuki}},\
  }\bibfield  {title} {\bibinfo {title} {Inelastic x-ray scattering studies of
  electronic excitations},\ }\href {https://doi.org/10.7566/JPSJ.82.021015}
  {\bibfield  {journal} {\bibinfo  {journal} {J. Phys. Soc. Jpn.}\ }\textbf
  {\bibinfo {volume} {82}},\ \bibinfo {pages} {021015} (\bibinfo {year}
  {2013}{\natexlab{a}})}\BibitemShut {NoStop}%
\bibitem [{\citenamefont {Ishii}\ \emph
  {et~al.}(2013{\natexlab{b}})\citenamefont {Ishii}, \citenamefont {Jarrige},
  \citenamefont {Yoshida}, \citenamefont {Ikeuchi}, \citenamefont {Inami},
  \citenamefont {Murakami},\ and\ \citenamefont
  {Mizuki}}]{Ishii.K_etal.J.-Electron.-Spectrosc.2013}%
  \BibitemOpen
  \bibfield  {author} {\bibinfo {author} {\bibfnamefont {K.}~\bibnamefont
  {Ishii}}, \bibinfo {author} {\bibfnamefont {I.}~\bibnamefont {Jarrige}},
  \bibinfo {author} {\bibfnamefont {M.}~\bibnamefont {Yoshida}}, \bibinfo
  {author} {\bibfnamefont {K.}~\bibnamefont {Ikeuchi}}, \bibinfo {author}
  {\bibfnamefont {T.}~\bibnamefont {Inami}}, \bibinfo {author} {\bibfnamefont
  {Y.}~\bibnamefont {Murakami}},\ and\ \bibinfo {author} {\bibfnamefont
  {J.}~\bibnamefont {Mizuki}},\ }\bibfield  {title} {\bibinfo {title}
  {Instrumental upgrades of the rixs spectrometer at BL11XU at SPring-8},\
  }\href {https://doi.org/https://doi.org/10.1016/j.elspec.2012.12.003}
  {\bibfield  {journal} {\bibinfo  {journal} {J. Electron. Spectrosc.}\
  }\textbf {\bibinfo {volume} {188}},\ \bibinfo {pages} {127 } (\bibinfo {year}
  {2013}{\natexlab{b}})}\BibitemShut {NoStop}%
\bibitem [{\citenamefont {Gretarsson}\ \emph {et~al.}(2013)\citenamefont
  {Gretarsson}, \citenamefont {Clancy}, \citenamefont {Liu}, \citenamefont
  {Hill}, \citenamefont {Bozin}, \citenamefont {Singh}, \citenamefont {Manni},
  \citenamefont {Gegenwart}, \citenamefont {Kim}, \citenamefont {Said},
  \citenamefont {Casa}, \citenamefont {Gog}, \citenamefont {Upton},
  \citenamefont {Kim}, \citenamefont {Yu}, \citenamefont {Katukuri},
  \citenamefont {Hozoi}, \citenamefont {van~den Brink},\ and\ \citenamefont
  {Kim}}]{Gretarsson.H_etal.Phys.-Rev.-Lett.2013}%
  \BibitemOpen
  \bibfield  {author} {\bibinfo {author} {\bibfnamefont {H.}~\bibnamefont
  {Gretarsson}}, \bibinfo {author} {\bibfnamefont {J.~P.}\ \bibnamefont
  {Clancy}}, \bibinfo {author} {\bibfnamefont {X.}~\bibnamefont {Liu}},
  \bibinfo {author} {\bibfnamefont {J.~P.}\ \bibnamefont {Hill}}, \bibinfo
  {author} {\bibfnamefont {E.}~\bibnamefont {Bozin}}, \bibinfo {author}
  {\bibfnamefont {Y.}~\bibnamefont {Singh}}, \bibinfo {author} {\bibfnamefont
  {S.}~\bibnamefont {Manni}}, \bibinfo {author} {\bibfnamefont
  {P.}~\bibnamefont {Gegenwart}}, \bibinfo {author} {\bibfnamefont
  {J.}~\bibnamefont {Kim}}, \bibinfo {author} {\bibfnamefont {A.~H.}\
  \bibnamefont {Said}}, \bibinfo {author} {\bibfnamefont {D.}~\bibnamefont
  {Casa}}, \bibinfo {author} {\bibfnamefont {T.}~\bibnamefont {Gog}}, \bibinfo
  {author} {\bibfnamefont {M.~H.}\ \bibnamefont {Upton}}, \bibinfo {author}
  {\bibfnamefont {H.-S.}\ \bibnamefont {Kim}}, \bibinfo {author} {\bibfnamefont
  {J.}~\bibnamefont {Yu}}, \bibinfo {author} {\bibfnamefont {V.~M.}\
  \bibnamefont {Katukuri}}, \bibinfo {author} {\bibfnamefont {L.}~\bibnamefont
  {Hozoi}}, \bibinfo {author} {\bibfnamefont {J.}~\bibnamefont {van~den
  Brink}},\ and\ \bibinfo {author} {\bibfnamefont {Y.-J.}\ \bibnamefont
  {Kim}},\ }\bibfield  {title} {\bibinfo {title} {Crystal-field splitting and
  correlation effect on the electronic structure of
  ${A}_{2}{\mathrm{IrO}}_{3}$},\ }\href
  {https://doi.org/10.1103/PhysRevLett.110.076402} {\bibfield  {journal}
  {\bibinfo  {journal} {Phys. Rev. Lett.}\ }\textbf {\bibinfo {volume} {110}},\
  \bibinfo {pages} {076402} (\bibinfo {year} {2013})}\BibitemShut {NoStop}%
\bibitem [{\citenamefont {Suzuki}\ \emph {et~al.}(2021)\citenamefont {Suzuki},
  \citenamefont {Liu}, \citenamefont {Bertinshaw}, \citenamefont {Ueda},
  \citenamefont {Kim}, \citenamefont {Laha}, \citenamefont {Weber},
  \citenamefont {Yang}, \citenamefont {Wang}, \citenamefont {Takahashi},
  \citenamefont {F{\"u}rsich}, \citenamefont {Minola}, \citenamefont {Lotsch},
  \citenamefont {Kim}, \citenamefont {Yava{\c s}}, \citenamefont {Daghofer},
  \citenamefont {Chaloupka}, \citenamefont {Khaliullin}, \citenamefont
  {Gretarsson},\ and\ \citenamefont {Keimer}}]{Suzuki.H_etal.Nat.-Commun.2021}%
  \BibitemOpen
  \bibfield  {author} {\bibinfo {author} {\bibfnamefont {H.}~\bibnamefont
  {Suzuki}}, \bibinfo {author} {\bibfnamefont {H.}~\bibnamefont {Liu}},
  \bibinfo {author} {\bibfnamefont {J.}~\bibnamefont {Bertinshaw}}, \bibinfo
  {author} {\bibfnamefont {K.}~\bibnamefont {Ueda}}, \bibinfo {author}
  {\bibfnamefont {H.}~\bibnamefont {Kim}}, \bibinfo {author} {\bibfnamefont
  {S.}~\bibnamefont {Laha}}, \bibinfo {author} {\bibfnamefont {D.}~\bibnamefont
  {Weber}}, \bibinfo {author} {\bibfnamefont {Z.}~\bibnamefont {Yang}},
  \bibinfo {author} {\bibfnamefont {L.}~\bibnamefont {Wang}}, \bibinfo {author}
  {\bibfnamefont {H.}~\bibnamefont {Takahashi}}, \bibinfo {author}
  {\bibfnamefont {K.}~\bibnamefont {F{\"u}rsich}}, \bibinfo {author}
  {\bibfnamefont {M.}~\bibnamefont {Minola}}, \bibinfo {author} {\bibfnamefont
  {B.~V.}\ \bibnamefont {Lotsch}}, \bibinfo {author} {\bibfnamefont {B.~J.}\
  \bibnamefont {Kim}}, \bibinfo {author} {\bibfnamefont {H.}~\bibnamefont
  {Yava{\c s}}}, \bibinfo {author} {\bibfnamefont {M.}~\bibnamefont
  {Daghofer}}, \bibinfo {author} {\bibfnamefont {J.}~\bibnamefont {Chaloupka}},
  \bibinfo {author} {\bibfnamefont {G.}~\bibnamefont {Khaliullin}}, \bibinfo
  {author} {\bibfnamefont {H.}~\bibnamefont {Gretarsson}},\ and\ \bibinfo
  {author} {\bibfnamefont {B.}~\bibnamefont {Keimer}},\ }\bibfield  {title}
  {\bibinfo {title} {Proximate ferromagnetic state in the Kitaev model material
  $\alpha$-RuCl$_3$},\ }\href {https://doi.org/10.1038/s41467-021-24722-4}
  {\bibfield  {journal} {\bibinfo  {journal} {Nat. Commun.}\ }\textbf {\bibinfo
  {volume} {12}},\ \bibinfo {pages} {4512} (\bibinfo {year}
  {2021})}\BibitemShut {NoStop}%
\bibitem [{\citenamefont {Lebert}\ \emph {et~al.}(2023)\citenamefont {Lebert},
  \citenamefont {Kim}, \citenamefont {Kim}, \citenamefont {Chun}, \citenamefont
  {Casa}, \citenamefont {Choi}, \citenamefont {Agrestini}, \citenamefont
  {Zhou}, \citenamefont {Garcia-Fernandez},\ and\ \citenamefont
  {Kim}}]{Lebert.B_etal.Phys.-Rev.-B2023}%
  \BibitemOpen
  \bibfield  {author} {\bibinfo {author} {\bibfnamefont {B.~W.}\ \bibnamefont
  {Lebert}}, \bibinfo {author} {\bibfnamefont {S.}~\bibnamefont {Kim}},
  \bibinfo {author} {\bibfnamefont {B.~H.}\ \bibnamefont {Kim}}, \bibinfo
  {author} {\bibfnamefont {S.~H.}\ \bibnamefont {Chun}}, \bibinfo {author}
  {\bibfnamefont {D.}~\bibnamefont {Casa}}, \bibinfo {author} {\bibfnamefont
  {J.}~\bibnamefont {Choi}}, \bibinfo {author} {\bibfnamefont {S.}~\bibnamefont
  {Agrestini}}, \bibinfo {author} {\bibfnamefont {K.}~\bibnamefont {Zhou}},
  \bibinfo {author} {\bibfnamefont {M.}~\bibnamefont {Garcia-Fernandez}},\ and\
  \bibinfo {author} {\bibfnamefont {Y.-J.}\ \bibnamefont {Kim}},\ }\bibfield
  {title} {\bibinfo {title} {Nonlocal features of the spin-orbit exciton in
  Kitaev materials},\ }\href {https://doi.org/10.1103/PhysRevB.108.155122}
  {\bibfield  {journal} {\bibinfo  {journal} {Phys. Rev. B}\ }\textbf {\bibinfo
  {volume} {108}},\ \bibinfo {pages} {155122} (\bibinfo {year}
  {2023})}\BibitemShut {NoStop}%
\bibitem [{\citenamefont {Gretarsson}\ \emph {et~al.}(2024)\citenamefont
  {Gretarsson}, \citenamefont {Fujihara}, \citenamefont {Sato}, \citenamefont
  {Gotou}, \citenamefont {Imai}, \citenamefont {Ohgushi}, \citenamefont
  {Keimer},\ and\ \citenamefont {Suzuki}}]{Gretarsson.H_etal.Phys.-Rev.-B2024}%
  \BibitemOpen
  \bibfield  {author} {\bibinfo {author} {\bibfnamefont {H.}~\bibnamefont
  {Gretarsson}}, \bibinfo {author} {\bibfnamefont {H.}~\bibnamefont
  {Fujihara}}, \bibinfo {author} {\bibfnamefont {F.}~\bibnamefont {Sato}},
  \bibinfo {author} {\bibfnamefont {H.}~\bibnamefont {Gotou}}, \bibinfo
  {author} {\bibfnamefont {Y.}~\bibnamefont {Imai}}, \bibinfo {author}
  {\bibfnamefont {K.}~\bibnamefont {Ohgushi}}, \bibinfo {author} {\bibfnamefont
  {B.}~\bibnamefont {Keimer}},\ and\ \bibinfo {author} {\bibfnamefont
  {H.}~\bibnamefont {Suzuki}},\ }\bibfield  {title} {\bibinfo {title}
  {$J$=$\frac{1}{2}$ pseudospins and $d\text{\ensuremath{-}}p$ hybridization in
  the Kitaev spin liquid candidates $\mathrm{Ru}{X}_{3}$ ($X$ = Cl, Br, I)},\
  }\href {https://doi.org/10.1103/PhysRevB.109.L180413} {\bibfield  {journal}
  {\bibinfo  {journal} {Phys. Rev. B}\ }\textbf {\bibinfo {volume} {109}},\
  \bibinfo {pages} {L180413} (\bibinfo {year} {2024})}\BibitemShut {NoStop}%
\bibitem [{\citenamefont {Takahashi}\ \emph {et~al.}(2021)\citenamefont
  {Takahashi}, \citenamefont {Suzuki}, \citenamefont {Bertinshaw},
  \citenamefont {Bette}, \citenamefont {M\"uhle}, \citenamefont {Nuss},
  \citenamefont {Dinnebier}, \citenamefont {Yaresko}, \citenamefont
  {Khaliullin}, \citenamefont {Gretarsson}, \citenamefont {Takayama},
  \citenamefont {Takagi},\ and\ \citenamefont
  {Keimer}}]{Takahashi.H_etal.Phys.-Rev.-Lett.2021}%
  \BibitemOpen
  \bibfield  {author} {\bibinfo {author} {\bibfnamefont {H.}~\bibnamefont
  {Takahashi}}, \bibinfo {author} {\bibfnamefont {H.}~\bibnamefont {Suzuki}},
  \bibinfo {author} {\bibfnamefont {J.}~\bibnamefont {Bertinshaw}}, \bibinfo
  {author} {\bibfnamefont {S.}~\bibnamefont {Bette}}, \bibinfo {author}
  {\bibfnamefont {C.}~\bibnamefont {M\"uhle}}, \bibinfo {author} {\bibfnamefont
  {J.}~\bibnamefont {Nuss}}, \bibinfo {author} {\bibfnamefont {R.}~\bibnamefont
  {Dinnebier}}, \bibinfo {author} {\bibfnamefont {A.}~\bibnamefont {Yaresko}},
  \bibinfo {author} {\bibfnamefont {G.}~\bibnamefont {Khaliullin}}, \bibinfo
  {author} {\bibfnamefont {H.}~\bibnamefont {Gretarsson}}, \bibinfo {author}
  {\bibfnamefont {T.}~\bibnamefont {Takayama}}, \bibinfo {author}
  {\bibfnamefont {H.}~\bibnamefont {Takagi}},\ and\ \bibinfo {author}
  {\bibfnamefont {B.}~\bibnamefont {Keimer}},\ }\bibfield  {title} {\bibinfo
  {title} {Nonmagnetic $J=0$ state and spin-orbit excitations in
  ${\mathrm{K}}_{2}{\mathrm{RuCl}}_{6}$},\ }\href
  {https://doi.org/10.1103/PhysRevLett.127.227201} {\bibfield  {journal}
  {\bibinfo  {journal} {Phys. Rev. Lett.}\ }\textbf {\bibinfo {volume} {127}},\
  \bibinfo {pages} {227201} (\bibinfo {year} {2021})}\BibitemShut {NoStop}%
\bibitem [{\citenamefont {Sugano}\ \emph {et~al.}(1970)\citenamefont {Sugano},
  \citenamefont {Tanabe},\ and\ \citenamefont {Kamimura}}]{Sugano.S_etal.1970}%
  \BibitemOpen
  \bibfield  {author} {\bibinfo {author} {\bibfnamefont {S.}~\bibnamefont
  {Sugano}}, \bibinfo {author} {\bibfnamefont {Y.}~\bibnamefont {Tanabe}},\
  and\ \bibinfo {author} {\bibfnamefont {H.}~\bibnamefont {Kamimura}},\
  }\href@noop {} {\emph {\bibinfo {title} {Multiplets of transition-metal ions
  in crystals}}}\ (\bibinfo  {publisher} {Academic Press},\ \bibinfo {address}
  {New York},\ \bibinfo {year} {1970})\BibitemShut {NoStop}%
\bibitem [{\citenamefont {Georges}\ \emph {et~al.}(2013)\citenamefont
  {Georges}, \citenamefont {Medici},\ and\ \citenamefont
  {Mravlje}}]{Georges.A_etal.Annu.-Rev.-Condens.-Matter-Phys.2013}%
  \BibitemOpen
  \bibfield  {author} {\bibinfo {author} {\bibfnamefont {A.}~\bibnamefont
  {Georges}}, \bibinfo {author} {\bibfnamefont {L.~d.}\ \bibnamefont
  {Medici}},\ and\ \bibinfo {author} {\bibfnamefont {J.}~\bibnamefont
  {Mravlje}},\ }\bibfield  {title} {\bibinfo {title} {Strong correlations from
  Hund's coupling},\ }\href
  {https://doi.org/10.1146/annurev-conmatphys-020911-125045} {\bibfield
  {journal} {\bibinfo  {journal} {Annu. Rev. Condens. Matter Phys.}\ }\textbf
  {\bibinfo {volume} {4}},\ \bibinfo {pages} {137} (\bibinfo {year}
  {2013})}\BibitemShut {NoStop}%
\bibitem [{\citenamefont {Chaloupka}\ and\ \citenamefont
  {Khaliullin}(2016)}]{Chaloupka.J_etal.Phys.-Rev.-B2016}%
  \BibitemOpen
  \bibfield  {author} {\bibinfo {author} {\bibfnamefont {J.}\
  \bibnamefont {Chaloupka}}\ and\ \bibinfo {author} {\bibfnamefont
  {G.}~\bibnamefont {Khaliullin}},\ }\bibfield  {title} {\bibinfo {title}
  {Magnetic anisotropy in the Kitaev model systems
  ${\mathrm{Na}}_{2}{\mathrm{IrO}}_{3}$ and ${\mathrm{RuCl}}_{3}$},\ }\href
  {https://doi.org/10.1103/PhysRevB.94.064435} {\bibfield  {journal} {\bibinfo
  {journal} {Phys. Rev. B}\ }\textbf {\bibinfo {volume} {94}},\ \bibinfo
  {pages} {064435} (\bibinfo {year} {2016})}\BibitemShut {NoStop}%
\bibitem [{\citenamefont {Abragam}\ and\ \citenamefont
  {Bleaney}(1970)}]{Abragam.A_etal.1970}%
  \BibitemOpen
  \bibfield  {author} {\bibinfo {author} {\bibfnamefont {A.}~\bibnamefont
  {Abragam}}\ and\ \bibinfo {author} {\bibfnamefont {B.}~\bibnamefont
  {Bleaney}},\ }\href@noop {} {\emph {\bibinfo {title} {Electron Paramagnetic
  Resonance of Transition Ions}}}\ (\bibinfo  {publisher} {Clarendon Press},\
  \bibinfo {address} {Oxford},\ \bibinfo {year} {1970})\BibitemShut {NoStop}%
\bibitem [{Zen()}]{Zenodo}%
  \BibitemOpen
  \href@noop {} {\bibinfo {title} {Url}}\BibitemShut {NoStop}%
\end{thebibliography}%

\end{document}